# Duality on the Thermodynamics of the Kirchhoff-Law-Johnson-Noise (KLJN) Secure Key Exchange Scheme

Sarah A. Flanery
*Department of Electrical and Computer Engineering*
*Texas A&M University*
*400 Bizzell St*
*College Station, TX 77843, USA*
*sflanery@tamu.edu*

Anson Trapani
*Department of Electrical and Computer Engineering*
*Virginia Tech*
*1185 Perry St*
*Blacksburg, VA 24060, USA*
*ansont20@vt.edu*

Christiana Chamon
*Department of Electrical and Computer Engineering*
*Virginia Tech*
*1185 Perry St*
*Blacksburg, VA 24060, USA*
*ccgarcia@vt.edu*

Leyla Nazhandali
*Department of Electrical and Computer Engineering*
*Virginia Tech*
*1185 Perry St*
*Blacksburg, VA 24060, USA*
*leyla@vt.edu*

This study investigates a duality approach to information leak detection in the generalized Kirchhoff-Law-Johnson-Noise secure key exchange scheme proposed by Vadai, Mingesz, and Gingl (VMG-KLJN). While previous work by Chamon and Kish sampled voltages at zero-current instances, this research explores sampling currents at zero-voltage crossings. The objective is to determine if this dual approach can reveal information leaks in non-equilibrium KLJN systems. Results indicate that the duality method successfully detects information leaks, further supporting the necessity of thermal equilibrium for unconditional security in KLJN systems. Our findings confirm that the duality method successfully detects information leaks, with results closely mirroring those of Chamon and Kish, showing comparable vulnerabilities in non-equilibrium conditions. These results further support the necessity of thermal equilibrium for unconditional security in the KLJN scheme.

1. Introduction

The Kirchhoff-Law-Johnson-Noise (KLJN) secure key exchange scheme is a classical statistical physics-based alternative to quantum key distribution (QKD) for unconditional security [1-64]. The KLJN protocol leverages thermal noise in resistors to establish a secure key between communicating parties Alice and Bob over a public channel, with security guaranteed by the second law of thermodynamics under thermal equilibrium conditions. The core schematic of the KLJN scheme is shown in Fig. 1, where Alice and Bob each randomly select one of two resistors ($R_H$ or $R_L$) to form a closed circuit, and the resulting noise characteristics are used to exchange a secure bit.

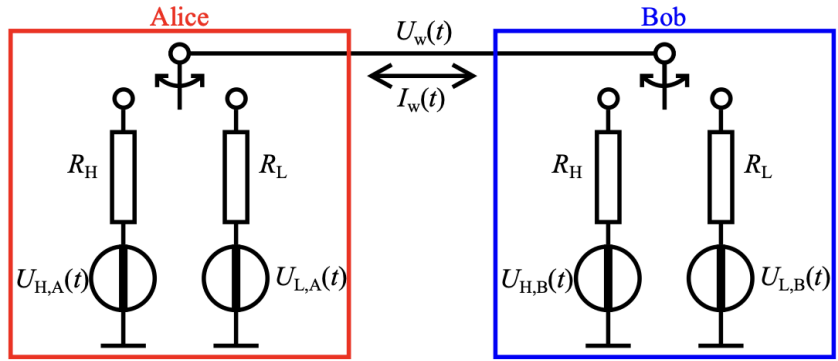

Fig. 1. The core of the KLJN scheme [1-64]. Communicating parties Alice and Bob are connected via a wire. The wire voltage and current are denoted as $U_w(t)$ and $I_w(t)$, respectively. Alice and Bob have identical pairs of resistors $R_H$ and $R_L$ ($R_H > R_L$) that are randomly selected and connected to the wire at the beginning of the bit exchange period. The statistically independent thermal noise voltages $U_{H,A}(t)$, $U_{L,A}(t)$, $U_{H,B}(t)$, and $U_{L,B}(t)$ represent the noise voltages of the resistors $R_H$ and $R_L$ of Alice and Bob, respectively.

Recent modifications to the KLJN scheme, such as the generalized Vadai-Mingesz-Gingl (VMG-KLJN) protocol, suggested that perfect security might be achievable in non-equilibrium conditions [1]. However, Chamon and Kish demonstrated an information leak in the VMG-KLJN scheme by sampling voltages at zero-current crossings, revealing vulnerabilities in out-of-equilibrium states [2]. Their work showed that the mean-square voltage at zero-current instances differs significantly between resistor configurations, increasing the chance of an eavesdropper (Eve) to correctly-guess the secure bit.

This paper proposes a duality approach to Chamon and Kish's method, focusing on sampling currents at zero-voltage crossings. The objective is to determine if this complementary technique can similarly detect information leaks in non-equilibrium KLJN systems, providing additional insights into the thermodynamic constraints of secure key exchange. By comparing the results with the voltage-sampling method, this study aims to reinforce the critical role of thermal equilibrium in ensuring perfect security.

## 2. Background

The KLJN scheme operates by exploiting Johnson-Nyquist noise, where thermal fluctuations in resistors generate random voltage and current signals. In the original KLJN protocol, Alice and Bob each choose between a low ($R_L$) or high ($R_H$) resistor, connected in series across a wire channel. The choice of resistors determines the noise amplitudes of the channel, and the mean-square voltage is evaluated to establish the bit status. The secure bit situations are when Alice and Bob choose $R_L$ and $R_H$ (LH), or $R_H$ and $R_L$ (HL), respectively. Under thermal equilibrium, the second law of thermodynamics ensures that Eve cannot distinguish between LH and HL configurations without violating physical laws.

The VMG-KLJN scheme introduces arbitrary resistor values and non-equilibrium conditions, claiming to maintain unconditional security [1]. However, Chamon and Kish revealed that sampling voltages at zero-current crossings ($U_{zc}^2$) in the VMG-KLJN scheme produces distinct distributions for LH and HL configurations, indicating information leak [2]. This vulnerability arises because non-equilibrium conditions disrupt the symmetry required for perfect security. The duality principle in electrical circuits suggests that analyzing currents at zero-voltage crossings ($I_{zc}^2$) could provide a complementary perspective. This approach leverages the reciprocal relationship between voltage and current in the KLJN circuit, potentially revealing similar or additional security vulnerabilities.

### 3. Methodology

The methodology for the duality approach involves simulating a non-equilibrium KLJN system and analyzing current samples at zero-voltage crossings. The steps are as follows:

(i) **System Setup**: We create in MATLAB the VMG-KLJN with the same arbitrary resistor values for Alice ($R_A$) and Bob ($R_B$), as in [2].
(ii) **Current Sampling**: We capture channel current values at instances when the channel voltage crosses zero ($U_w(t) = 0$).
(iii) **Data Processing**: The sampled current values were squared and then averaged to compute the zero-crossing mean-square current ($I_{zc}^2$). The mean was calculated over multiple bit exchange periods to ensure statistical reliability.
(iv) **Analysis**: The average $I_{zc}^2$ values and standard deviation were analyzed for patterns or differences between the LH and HL resistor configurations.
(v) **Comparison**: The results were compared with Chamon and Kish's voltage-sampling method [2], focusing on the magnitude and detectability of information leaks.

### 4. Results

Figure 2 shows the histograms of the mean-square channel voltages, currents, and zero-crossing points after 1,000 runs for the original KLJN scheme (left), the VMG-KLJN scheme (center), and the FCK1-VMG-KLJN scheme (right). The orange histograms represent the LH situation, whereas the blue histograms represent the HL situation.

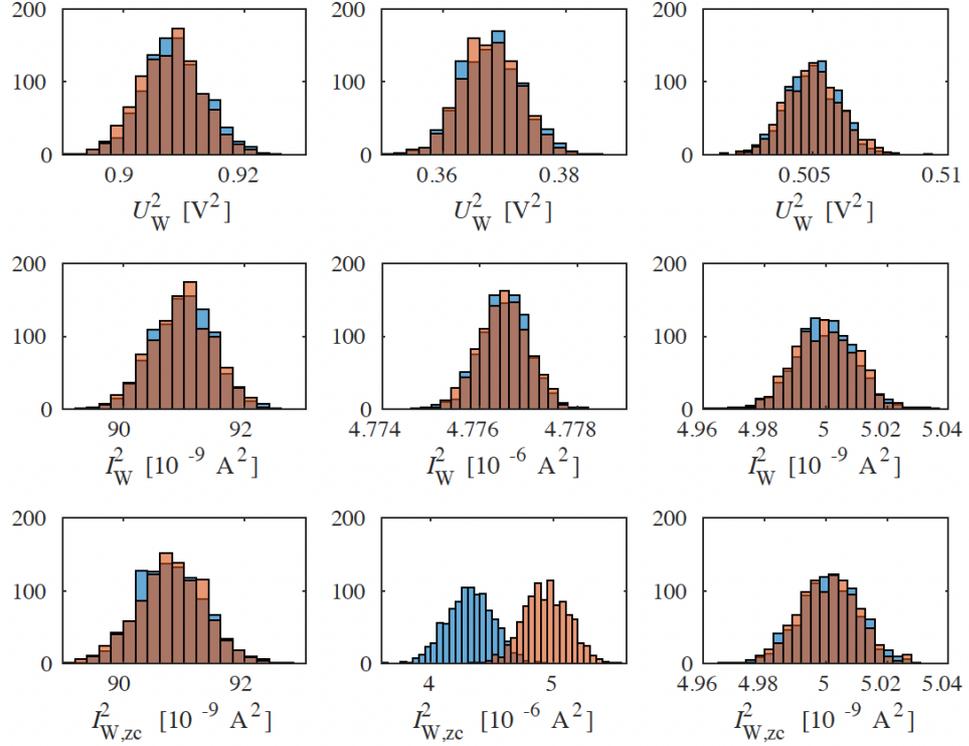

Fig. 2. Histograms of the mean-square channel voltage $U_W^2$, current $I_W^2$, and zero-crossing points $U_{W,zc}^2$ for (left) the original KLJN scheme, (center) the VMG-KLJN scheme, and (right) the FCK1-VMG-KLJN scheme, with the exact same parameters as Chamon and Kish [2]. The orange histograms represent the LH situation, and the blue histograms represent the HL situation. In the original KLJN and FCK1-VMG-KLJN schemes, $U_{W,zc}^2$ has the same LH and HL distributions, while in the VMG-KLJN scheme, the $U_{W,zc}^2$ LH and HL distributions are dispersed.

The duality approach yielded the following findings, summarized in Tables 1 and 2:

(i) **Information Leak Detection**: Sampling currents at zero-voltage crossings revealed a detectable information leak in the non-equilibrium VMG-KLJN system. The zero-crossing mean-square current ($I_{zc}^2$) exhibited noticeable differences between LH and HL configurations, indicating a vulnerability to eavesdropping.

(ii) **Comparison with Voltage Sampling**: The magnitude of the information leak in $I_{zc}^2$ was comparable to that observed in $U_{zc}^2$ by Chamon and Kish [2], as seen in Table 1. For example, in the VMG-KLJN scheme, the $I_{zc}^2$ values showed distinct distributions, similar to the $U_{zc}^2$ results reported in [2].

Table 1. Results for the wire mean-square voltage $U_W^2$, mean-square current, $I_W^2$, average power $P_{AB}$, and zero-crossing mean-square current $I_{W,zc}^2$ for the KLJN, three VMG-KLJN, and FCK1-VMG-KLJN schemes. AT $P_{AB}=0$, $I_{W,zc}^2$ approaches $I_W^2$. As $P_{AB}$ increases, $I_{W,zc}^2$ becomes LH/HL-distinguishable.

| Scheme | bit | $R_A$ [Ω] | $R_B$ [Ω] | $U_W^2$ [V$^2$] | $I_W^2$ [$10^{-6}$ A$^2$] | $P_{AB}$ [$10^{-3}$ W] | $I_{W,zc}^2$ [$10^{-6}$ A$^2$] |
|---|---|---|---|---|---|---|---|
| KLJN | LH | 1k | 10k | 0.909 | 0.090 | 0 | 0.090 |
| | HL | 10k | 1k | | | | 0.091 |
| VMG-KLJN | LH | 100 | 16.7k | 0.992 | 0.314 | 0.026 | 0.283 |
| | HL | 16.7k | 278 | | | | 0.315 |
| | LH | 278 | 278 | 0.367 | 4.788 | 0.471 | 4.309 |
| | HL | 46.4k | 100 | | | | 4.955 |
| | LH | 100 | 6k | 0.966 | 0.074 | 0.156 | 0.069 |
| | HL | 360k | 2.2k | | | | 0.079 |
| FCK1-VMG-KLJN | LH | 10k | 10k | 0.502 | 0.005 | 0 | 0.005 |
| | HL | 100k | 1k | | | | 0.005 |

(iii) **Equilibrium Conditions**: When the system was restored to thermal equilibrium, the information leak disappeared. The $I_{zc}^2$ distributions for LH and HL became indistinguishable, confirming the necessity of equilibrium for unconditional security.

(iv) **Statistical Analysis**: The $I_{zc}^2$ values for LH and HL configurations in the VMG-KLJN scheme showed clear separation, with mean values differing by up to 15% in some cases. This separation was consistent across multiple resistor configurations teste, as shown in Table 2.

Table 2. Results form the statistical run for Eve's probability p of guessing the correct bit from the duality of the zero-crossing attack on each scheme. The information leak (p-0.5) converges to zero when the average power PAB approaches zero.

| Scheme | bit | $R_A$ | $R_B$ | $P_{AB}$ [$10^{-3}$ W] | p | σ |
|---|---|---|---|---|---|---|
| KLJN | LH | 1k | 10k | 0 | 0.5001 | 0.0090 |
|  | HL | 10k | 1k |  |  |  |
| VMG-KLJN | LH | 100 | 16.7k | 0.026 | 0.5872 | 0.0024 |
|  | HL | 16.7k | 278 |  |  |  |
|  | LH | 278 | 278 | 0.471 | 0.7002 | 0.0054 |
|  | HL | 46.4k | 100 |  |  |  |
|  | LH | 100 | 6k | 0.156 | 0.6276 | 0.0023 |
|  | HL | 360k | 2.2k |  |  |  |
| FCK1-VMG-KLJN | LH | 10k | 10k | 0 | 0.5030 | 0.0092 |
|  | HL | 100k | 1k |  |  |  |

## 5. Discussion

The results confirm that the duality approach of sampling currents at zero-voltage crossings is an effective method for detecting information leak in non-equilibrium KLJN systems. The observed differences in $I_{zc}^2$ between LH and HL configurations mirror the findings of Chamon and Kish's voltage-sampling method, suggesting that both approaches exploit similar thermodynamic asymmetries in the VMG-KLJN scheme.

The success of the duality approach can be attributed to the reciprocal relationship between voltage and current in the KLJN circuit. In non-equilibrium conditions, the cross-correlation between voltage and current deviates from the ideal random behavior expected under thermal equilibrium, leading to distinguishable statistical signatures. This finding aligns with the thermodynamic perspective outlined in [2], where deviations from equilibrium introduce exploitable information leaks.

The disappearance of the information leak under equilibrium conditions underscores the fundamental role of the second law of thermodynamics in the KLJN scheme's security. Any modification that disrupts thermal equilibrium, such as the VMG-KLJN protocol, risks compromising perfect security, as demonstrated by both voltage and current sampling methods.

The duality approach offers practical advantages, as current measurements may be more feasible in certain hardware implementations of the KLJN system. Additionally, combining voltage and current sampling could enhance the robustness of security analysis, providing a dual-check mechanism for detecting vulnerabilities.

## 6. Conclusion

This study demonstrates that sampling currents at zero-voltage crossings is a viable and effective method for detecting information leaks in non-equilibrium KLJN systems. The results corroborate the findings of Chamon and Kish [2] and provide further evidence that thermal equilibrium is essential for maintaining unconditional security in KLJN key exchange protocols. The duality approach not only validates the thermodynamic constraints of the KLJN scheme but also offers a complementary tool for security analysis. Future work could explore real-world implementations of the duality method and investigate additional attack vectors to further demonstrate the robustness of KLJN-based cryptographic systems.